\documentclass[epj]{svjour}
\usepackage{graphics}
\usepackage{epstopdf}

\begin{document}
\title{Resonance production from jet fragmentation}
\author{Christina Markert \inst{1,2} for the STAR Collaboration}                     % Do not remove
\institute{University of Texas, Austin, Texas 78712, USA \and \email{cmarkert@physics.utexas.edu}}%} 

\offprints{}          % Insert a name or remove this line
%
%\email{tone421@rcf.rhic.bnl.gov}
%
\date{Received: date / Revised version: date}
% \email{tone421@rcf.rhic.bnl.gov}
\abstract{
Short lived resonances are sensitive to the medium properties in heavy-ion collisions. Heavy hadrons have larger probability to be produced within the quark gluon plasma phase due to their short formation times. Therefore heavy mass resonances are more likely to be affected by the medium, and the identification of early produced resonances from jet fragmentation might be a viable option to study chirality. The high momentum resonances on the away-side of a triggered di-jet are likely to be the most modified by the partonic or early hadronic medium. We will discuss first results of triggered hadron-resonance correlations in Cu+Cu heavy ion collisions.
\PACS{
      {PACS-key}{discribing text of that key}   \and
      {PACS-key}{discribing text of that key}
     } % end of PACS codes
} %end of abstract
\maketitle
\section{Introduction}
\label{intro}

Chiral symmetry restoration is one of the fundamental features of QCD
at high temperatures and densities of nuclear matter which should occur
around the phase transition (0.8-1.2 $T_{\rm c}$) according to lattice QCD 
calculations \cite{lattice}. The signature of chiral symmetry  restoration could 
be mass shifts and/or broadenings of the hadronic resonance states inside
the medium created in a heavy ion collision.  Resonances from jet fragmentation can be produced early, depending on their hadronic formation times and might be medium modified through interactions with either the late partonic (1.2 $T_{\rm c}$) or early hadronic (0.8 $T_{\rm c}$) medium. To measure this effect, resonances also need to decay inside the medium.
The formation time of hadrons from jet fragmentation depends on their mass and momentum.  I will discuss the relevant momentum range for different resonances. Furthermore one
needs to select the resonances on the away-side of a surface biased (triggered) jet in order to maximize their in-medium interactions. This will be discussed in section \ref{resinjets}.

\section{Formation time}
\label{formationtime}

\begin{figure}[h]
\resizebox{0.47\textwidth}{!}{%
  \includegraphics{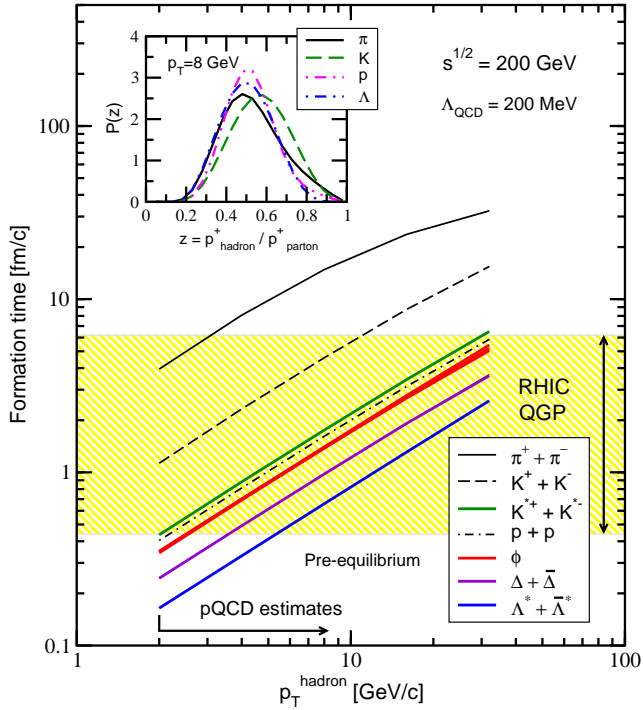}
}
%\vspace{5cm}       % Give the correct figure height in cm
\caption{Transverse momentum
dependence of leading hadron formation times at RHIC energies,
$\sqrt{s}=200$~GeV. Results for both 'stable' particles and
resonances are presented. The shaded area represents the estimated
partonic lifetime at RHIC. Insert shows the meson and baryon
distributions $P(z)$ at a fixed hadron momentum $p_T = 8$~GeV/c. The shaded area represents the estimated partonic lifetime at LHC using a Bjorken expansion.\cite{formnew}}
\label{rhic}       % Give a unique label
\end{figure}

\begin{figure}[h]
\resizebox{0.47\textwidth}{!}{%
  \includegraphics{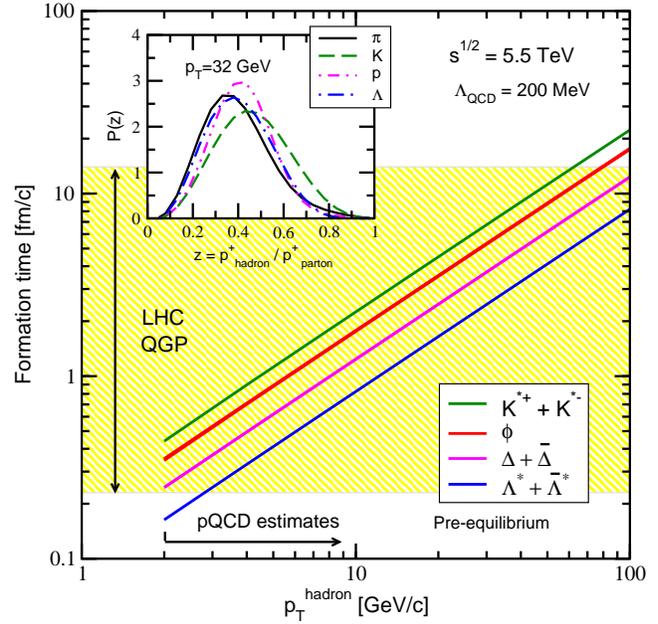}
}
%\vspace{5cm}       % Give the correct figure height in cm
\caption{Transverse momentum
dependence of heavy resonance formation times at LHC  energies,
$\sqrt{s}=5.5$~TeV. Insert shows the meson and baryon
distributions $P(z)$ at a fixed hadron momentum  $p_T = 32$~GeV/c.
The shaded area represents the estimated
partonic lifetime at LHC using a Bjorken expansion
\cite{formnew}.}
\label{lhc}       % Give a unique label
\end{figure}

The formation time of hadrons and resonances from jet fragmentation can be calculated  in a quantum 
mechanical approach \cite{vitev,formnew}. An alternative approach, based on string fragmentation \cite{falter}, arrives at a similar conclusion for heavy mass hadrons, light quark objects. 
A heavier hadron mass leads to a shorter formation time while a higher momentum causes a longer formation time. In the theoretical calculations \cite{formnew} the probability of high momentum heavy hadron (or resonance) formation in the partonic medium is finite. Therefore we expect
the resonance to be medium modified, after its formation, through parton-resonance interactions
in the so called mixed degree of freedom phase of partons and hadrons. The proper momentum range needs to be defined by a lower and an upper limit to ensure the formation and decay of the resonance inside the chiral medium. If the width of the resonances is broadened in the medium their lifetimes is shortened which will increase the probability of the resonances decaying inside the medium.  This would also result in a broader signal width in the invariant mass spectrum which decreases the statistical significance of the signal. 
Figure~\ref{rhic} shows the linear dependence of the average formation time on the transverse momentum of the hadron (or resonance) integrated over the fragmentation distribution of all produced jets at RHIC. The yellow shaded area is the lifetime of the QGP calculated by assuming a longitudinal Bjorken expansion with $T_c \approx 180$~MeV for the critical temperature. The estimated lifetime at RHIC is $\tau_{QGP} = 6.2$~fm/c and at the LHC is $\tau_{QGP} = 14$~fm/c based on an longitudinal Bjorken expansion \cite{formnew}. The RHIC QGP lifetime is in agreement with the partonic lifetime, derived from the resonance suppression in the hadronic medium  \cite{resostar,tor01,raf01,raf02,mar02} and the pion HBT source lifetime measurement  ($\Delta\tau=5-12$~fm/c) ~\cite{nig05}. Figure~\ref{lhc} shows the corresponding linear dependence of the mean formation time on the transverse momentum of the hadron (or resonance) for the LHC.

\section{Resonances in jets}
\label{resinjets}

In order to extract the chirally restored resonances from the partonic and early hadronic 
medium, a correlation analysis using a jet or hadron triggered high momentum resonance is suggested.
This also ensures that the resonance signal is largely unaffected by the late hadronic medium where
regeneration of resonances will dilute it. UrQMD 
calculations suggest that the regeneration of resonances is predominant in the low
momentum region (p$_{\rm T} <$ 2~GeV/c) \cite{urqmd}, the study of
chiral symmetry restoration is only
suitable in the momentum region of p$_{\rm T} >$ 2~GeV/c.
High momentum resonances are less affected by the re-scattering and
regeneration in the hadronic phase because the system moves with a smaller collective velocity than the probe, and thus the high momentum decay particles can escape.
Resonances from jets where the hard scattering occurs near the
surface are in particular unaffected by the medium. Thus the hadronic resonances from the same side jet 
in a triggered di-jet event will serve as the \`~de facto~\`  in-vacuum reference for the medium modification measurement. The away-side distribution of jet resonances is
selected to study chirality. Resonances in this distribution are likely to experience a
longer path length in the medium. However their momenta should be large
enough to be unaffected by re-scattering and regeneration in the subsequently
produced hadronic medium. This idea requires the resonances to be formed
earlier than the bulk hadrons, such that hadronic resonances can traverse the
partonic medium.

\section{Resonance correlations in STAR}

\begin{figure}[h]
\resizebox{0.5\textwidth}{!}{%
  \includegraphics{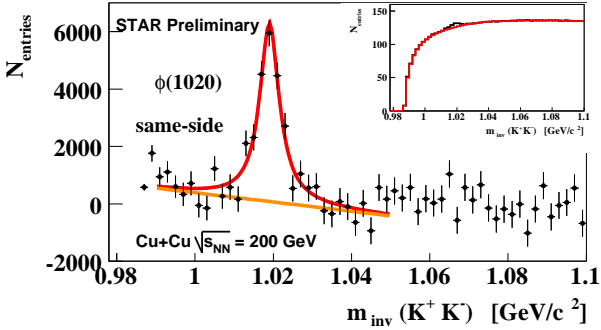}}
%\vspace{5cm}       % Give the correct figure height in cm
\caption{Invariant mass distributions of $\phi(1020) \rightarrow K^{+}+K^{-}$  ($p_{\rm T}$=(1-2) GeV/c ) for the in jet plane angular correlation (same-side) $\Delta\phi$=[ -1/18 $\pi$ , 3/18 $\pi$] with respect to a $p_{\rm T}$=(3-10) GeV/c hadron trigger in Cu+Cu collisions
at $\sqrt{s_{\rm NN}}=200\;\rm GeV$ before (inset) and after
mixed-event background subtraction.}
\label{invmasssame}       % Give a unique label
\end{figure}

\begin{figure}[h]
\resizebox{0.5\textwidth}{!}{%
  \includegraphics{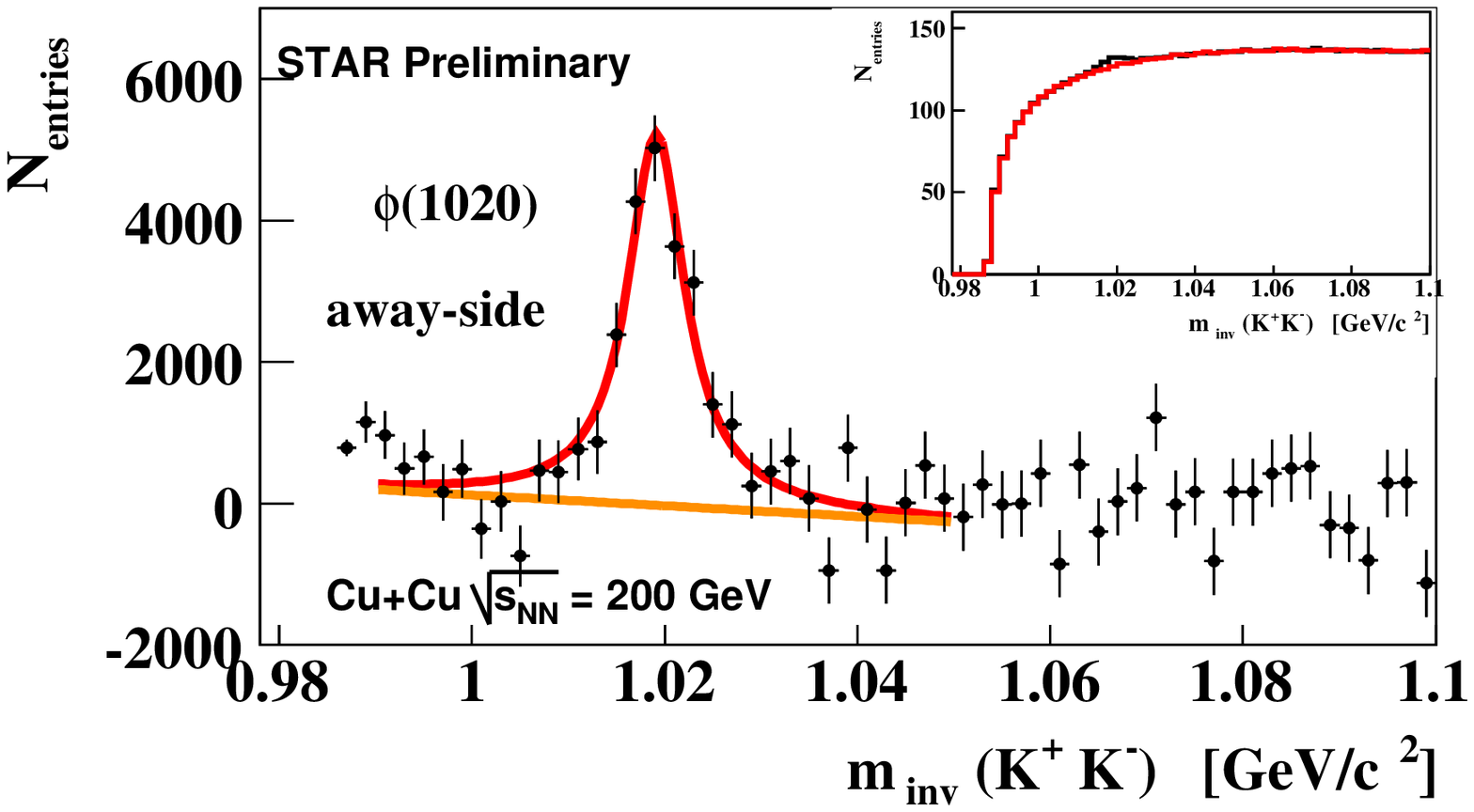}}
%\vspace{5cm}       % Give the correct figure height in cm
\caption{Invariant mass distributions of $\phi(1020) \rightarrow K^{+}+K^{-}$  ($p_{\rm T}$=(1-2) GeV/c ) for the in jet plane angular correlation (away-side) $\Delta\phi$= [19/18 $\pi$ , 23/18 $\pi$] with respect to a $p_{\rm T}$=(3-10) GeV/c hadron trigger in Cu+Cu collisions
at $\sqrt{s_{\rm NN}}=200\;\rm GeV$ before (inset) and after
mixed-event background subtraction.}
\label{invmassaway}       % Give a unique label
\end{figure}

A first attempt to study the high momentum charged hadron-$\phi$(1020) resonance correlation using the STAR detector was shown last year \cite{marjet}. The new analysis presented in these proceedings  
is using 2.33 M min bias Cu+Cu events at $\sqrt{s_{NN}}$ = 200 GeV. In the correlation analysis the trigger hadron is required to have a $p_{\rm T}$$>$ 3~GeV/c (2.58 M trigger) and an associated  $\phi$(1020) meson  with $p_{\rm T}$= 1-2~GeV/c. This momentum range was chosen to select mostly events including only one trigger hadron. In the future we would like to place a higher momentum cut on the trigger hadron to select the more energetic jets.
According to the formation time calculations one would prefer to select $\phi$(1020) ($p_{\rm T}$= (1-2)~GeV/c) from jet fragmentation between $p_{\rm T}$= 3-10~GeV/c, but due to the lack of statistics in the available STAR dataset we are limited to a smaller momentum range and thus this study serves more as a proof of principle.  
The $\Delta\phi$ distribution between the trigger hadron and the associated  $\phi$(1020) meson  is obtained by determining a yield from the invariant mass signal independently in each $\Delta\phi$ bin.
As an example, Figure~\ref{invmasssame} and 
Figure~\ref{invmassaway} show the invariant mass distribution and extracted $\phi$(1020) meson yield for the $\Delta\phi$=[-1/18 $\pi$ , 3/18 $\pi$]  (same-side) peak bin and   $\Delta\phi $=[19/18 $\pi$ , 23/18 $\pi$] (away-side) peak bin. The signal is shown after mixed event background subtraction. The uncorrelated background is generated with Kaon pairs where the two Kaons are from different evens. The events however belong to the same event category described by the vertex position of the reaction, the reaction plane and the multiplicity.

 %  p3           1.01904e+00   2.22765e-04  -2.23227e-04   2.22765e-04
  %  p4           6.00977e-03   6.71080e-04  -6.41511e-04   7.22433e-04
% entries 22034

%
%
%  p3           1.01914e+00   2.93142e-04  -2.89787e-04   2.97591e-04
%  p4           7.17752e-03   8.71258e-04  -8.44682e-04   9.44787e-04
% entries 20871

 %number of jet events all: 2.3273e+06
 %number of jet trigger all: 2.58387e+06

  \begin{figure}[h]
\resizebox{0.5\textwidth}{!}{%
  \includegraphics{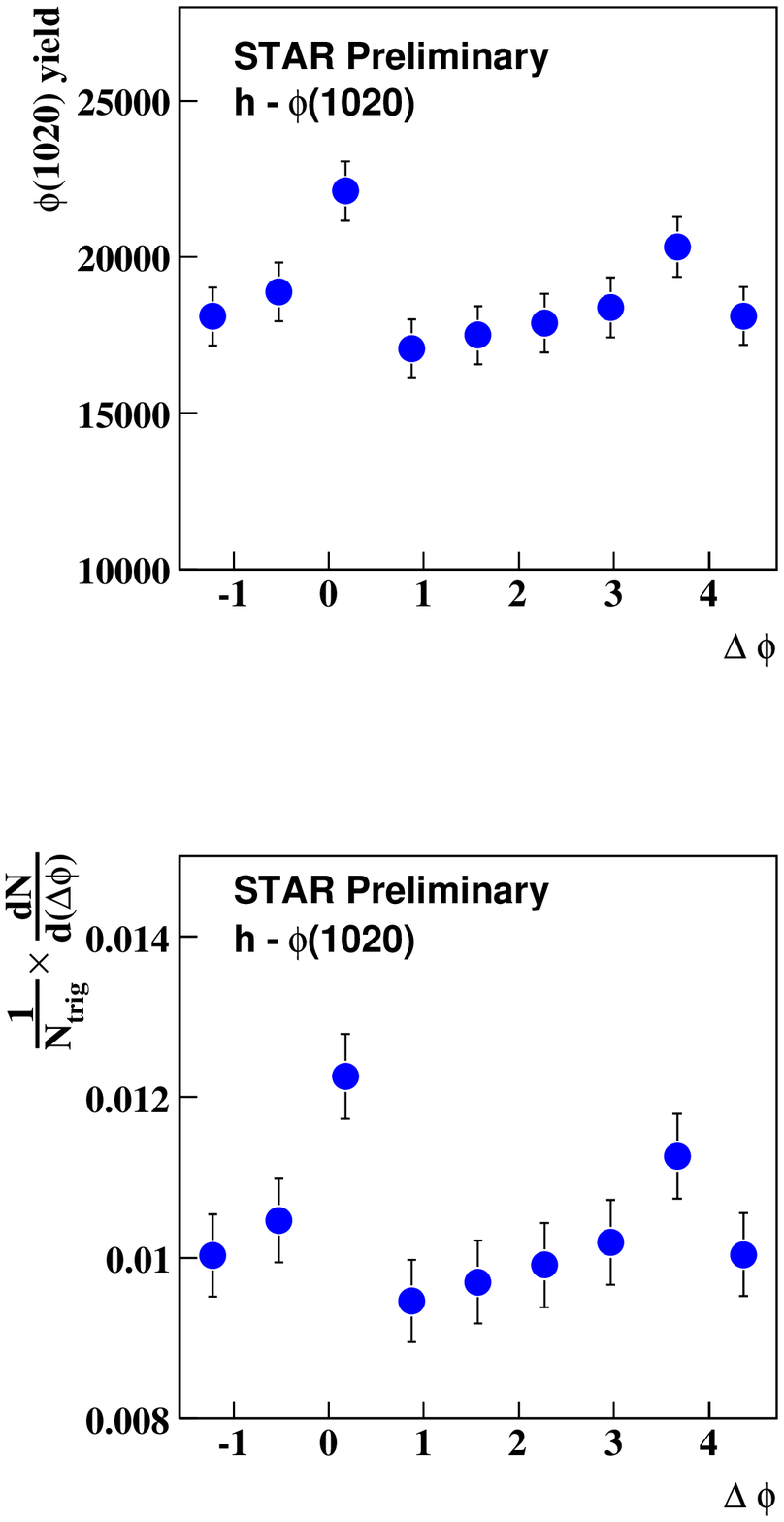}
}
%\vspace{5cm}       % Give the correct figure height in cm
\caption{Angular correlation of hadron-$\phi$(1020) resonance.
Hadron trigger p$_{\rm T}$ $>$ 3~GeV/c and associated $\phi$(1020)
 $p_{\rm T}$=(1-2) GeV/c. Upper:  $\phi$(1020) yield versus angle between  $\phi$(1020) and hadron trigger. Lower: Normalized by number of trigger hadrons. The errors are statistical only. The distribution is not corrected for elliptic flow v$_{2}$. The y-axis of the plots is zero suppressed.}
\label{deltaphi}       % Give a unique label
\end{figure}

The extracted yield for all $\Delta\phi$ angles is shown in Figure~\ref{deltaphi}. These data are not corrected for elliptic flow. The derived mass and width of the same-side signal (Fig~\ref{invmasssame}) is m = 1019.0 $\pm$ 0.3 MeV/c$^{2}$ and  $\Gamma$ = 6.0 $\pm$ 0.7 MeV/c$^{2}$ based on fit using a Breit-Wigner function combined with a linear background function. The derived mass and width of the away-side signal (Fig~\ref{invmassaway}) is m = 1019.1 $\pm$ 0.3 MeV/c$^{2}$ and  $\Gamma$ = 7.2 $\pm$ 0.9 MeV/c$^{2}$. An additional systematic error of 2 MeV/c$^{2}$ for the width depending on the fit range and normalisation of the background needs to be applied in all $\Delta\phi$ bins. The derived mass and width values are in agreement with the PDG values, when folded with the detector momentum resolution and energy loss in the detectors.

We are using the $\phi$(1020) as the associated particle
since its reconstructed mass spectrum has the largest significance
of all short lived resonances in STAR. However the lifetime of the
$\phi$(1020) is about 45 fm/c, which means that the majority will
decay outside of the medium. The 20M minimum bias Cu+Cu events analyzed
are not sufficient to place an effective high momentum cut on the $\phi$(1020) spectrum. The analyzed data show no evidence for mass shifts and width broadenings of resonances in the momentum range of  $p_{\rm T}$=(1-2) GeV/c, neither on the same side nor on the away side. This was to be expected since $\phi$(1020) mesons in this momentum range are most likely dominated by interactions in the late hadronic phase and will subsequently decay outside of the fireball due to the long lifetime of the $\phi$(1020). In addition the jet signal is sitting on a large background (S/B= 0.2). Therefore the invariant mass yield of the $\phi$(1020) resonance in this analysis is dominated by non-jet triggered entries. To enhance the resonances from jets a higher momentum range is required.

%%%% 

Presently, the momentum of
the kaon candidates for the $\phi$(1020) reconstruction is
restricted to p$_{\rm T}$ (0.2-1.1) GeV/c in order to achieve clean
particle identification in the TPC. In the future we will be able to
select the higher momentum $\phi$(1020)s with better significance by
using the additional Time of Flight (TOF) detector ~\cite{tof} which
enables us to identify kaons up to p = 1.5-2.0 GeV/c. \\
The requirement that the resonance not only needs to form but also decay inside the medium, in order to probe chiral symmetry in the boosted laboratory frame, drives the need for very short lived resonances in this analysis. 
Therefore K(892) or $\Delta$(1232) resonances are more suited than, for example, the $\phi$(1020).
Nevertheless any additional medium modification might broaden the width and  therefore
shorten the lifetime, so that a $\phi$(1020) might be still a good candidate \cite{haglin}.
The probability of the medium modified resonance decaying in the radially expanding plasma has only been qualitatively addressed at present. Calculations to determine the fraction of resonance decays in the medium are underway.

\section{Conclusions}
\label{conclusion}

Heavy hadronic resonances from jets are suitable to
study chiral symmetry restoration due to their short formation time. Theoretical calculations show that we expect short formation times of a few fm/c  for heavy mass hadrons up to  p$_{\rm T}$$=$10 GeV/c.
An analysis technique based on jet correlations to enrich the resonances
which pass through the medium has been proposed and also used to explore the phenomena in Cu+Cu collisions. Future detector upgrades such as the TOF detector for STAR are necessary to significantly reduce the background in the reconstruction of hadronically decaying high momentum resonances.

%
% BibTeX users please use
% \bibliographystyle{}
% \bibliography{}

\end{document}